\documentclass[sigconf, language=english]{acmart}

\AtBeginDocument{%
  \providecommand\BibTeX{{%
    \normalfont B\kern-0.5em{\scshape i\kern-0.25em b}\kern-0.8em\TeX}}}

\begin{document}

\title{O Contract, Where Art Thou?\\Contract Management as a SharePoint Oddity}

\author{Sasha Vtyurina}
\orcid{0000-0003-1501-3624}
\affiliation{%
  \institution{Zuva Inc}
  \country{Canada}
}

\author{Adam Roegiest}
\email{adam@zuva.ai}
\orcid{0000-0003-1265-8881}
\affiliation{%
  \institution{Zuva Inc}
  \country{Canada}
}



\begin{abstract}
For many legal operations teams, the management of the contracts and agreements that their
organization are negotiating or have been executed is an encompassing and time-consuming task. This has resulted in specialized tools for Contract Lifecycle Management (CLM) have grown steadily in demand over the last decade. Transitioning to such tools can itself be an arduous and costly process and so a logical step would be to augment existing storage solutions. In this paper, we present the analysis of 26 semi-structured interviews with legal operations professionals about their trials and tribulations with using Microsoft SharePoint for contract  management. We find that while there is promise, too much of what is needed to be successful requires more technical prowess than might be easily available to those empowered to put it in place.

\end{abstract}

\maketitle

\section{Introduction}

``O Contract, where art thou?'' is a, perhaps stylized, question that lawyers working at large organizations are likely to ask themselves or others on a not uncommon basis.
As business routinely deal with a wide assortment of contracts (e.g., lease agreements, non-disclosure agreements, supply agreements, employment agreements) in a varying set of processes that make it easy and likely to misplace (or perhaps, mis-save) a particular (version of a) contract. 
These processes range from the initial contract drafting, potentially from one of many templates, to lengthy negotiations with reviews by multiple individuals, to collecting final signatures, and finally to on-going obligation and requirement tracking (e.g., when to invoice, when to renew). 
Supporting such complex processes with a generally large number of people involved, both internal and external to the organization, means a great deal of effort can be spent on coordination among involved individuals around where versions of contracts are stored (or should be stored) and how they should be worked on.

This has resulted in the dramatic growth in a segment of software, called Contract Lifecycle Management (CLM), that seeks to encapsulate the entire lifecycle of a contract in a single system to remove much of the explicit coordination and heavy work to support contracts (Figure \ref{fig:lifecycle} depicts a high-level view of the contract lifecycle)~\cite{clm-market,gartner}.
These systems seek to take haphazard processes and ensconce them in a more structured and formal process that enables users to consistently negotiate and sign contracts without having to do nearly as much heavy lifting in the process.
The ``dark side'' of CLMs is that they usually require some form of migration of contracts from existing document repositories (e.g., Google Drive, Dropbox, Microsoft SharePoint) into the system and all the minute details that can require manual intervention (e.g., rebuilding versioning, collecting metadata).
Accordingly, there exists the potential for two barriers to form when looking at transitioning to a CLM: the actual price of a CLM and the cost of adopting the CLM itself. 
While the former cost can generally be mitigated (via negotiation) or appropriately budgeted for over time, the  latter cost can form a barrier to adoption since it invariably pulls people away from their day jobs to work that may not always have an apparent return on investment to executives or other stakeholders that can impede the process. 

Mitigating the human and effort costs in adopting CLMs can mean that organizations will seek to gain the benefits of CLMs through their existing tools and processes. 
For example, an organization may seek to leverage Google Workspace to create contract templates in Google Docs, use Google Docs to fill-in the templates, and then use an add-on to facilitate signatures via an e-signature application (e.g., Docusign\footnote{\url{https://www.docusign.com/}}, PandaDoc\footnote{\url{https://www.pandadoc.com/}}) that also put the signed version in a particular folder.
Based upon preliminary ad-hoc conversations conducted by other members of the team with 57 individuals, primarily more senior operations roles (e.g., executives), we observed that a Microsoft SharePoint was a common document repository for legal teams to store contracts and other agreements.
With this in mind, we sought to investigate how organizations use SharePoint for contract management and whether or not similar functionality to existing CLMs would be beneficial and whether such functionality would be easily implementable in SharePoint using other Microsoft tools (e.g., PowerAutomate\footnote{\url{https://powerautomate.microsoft.com/}}, PowerApps\footnote{\url{https://powerapps.microsoft.com/}}).




In this paper, we report on the results of 26 semi-structured interviews of individuals in a legal operations role (e.g., drafting, negotiation) whose organization uses or used to use Microsoft SharePoint as a document repository.
We find among our interviewees that many only focus on a few of the aspects of the contract lifecycle and that their needs using SharePoint are both inconsistent (i.e., exact overlap of needs is low) and that they all follow relatively bespoke processes to make SharePoint work for them. 
Many end-users have ad-hoc processes that have evolved to fill gaps and that they may prefer more established solutions to these problems rather than trying to determine best practices themselves.
Despite these desires, SharePoint does not make it easy for non-technical users to build workflows into their SharePoint sites via related products (e.g., PowerAutomate is accessible but perhaps not to a lawyer lacking familiarity with how to connect different services) nor is it ``turn key'' for them to buy existing solutions since they often require some level of setup that requires additional assistance (e.g., from their IT department).


\section{Background}
To an extent, contract lifecycle management can resemble the well studied known-item retrieval~\cite{dumais2003,10.1145/792550.792552,10.1145/3176349.3176392,10.1145/3406522.3446021} but there are several substantive differences. 
The biggest difference is that the known-item in this case may span several different ``documents'' due to amendments, renewals, and supplementary context (e.g., side letters specifying additional details outside the contract) which means that the "known" contract is not just a single item but could be one of many. 
Moreover, the most up-to-date information can change over time due to amendments and so finding the right documents can become problematic (e.g., finding an old amendment over the newest one).
In the ideal case, these details can be pulled out separately for more easy access and retrieval (e.g., returning the most recent version of a clause) but still requires some level of incorporation into systems and processes.

Many of the elements of contract (lifecycle) management resemble those of traditional enterprise search~\cite{hawking2004challenges,INR-053,kruschwitz20233,russell2011taxonomy,stocker2014,mukherjee2004} in that the documents being managed are internal to an organization
and that they are often retrieved to perform comparisons or other analyses. On the other hand, these documents are often not retrievable by employees at large but select individuals, usually lawyers or other sufficiently experienced employees, and are used, in part, to also govern the business (e.g., obligations to be met, payments to be made). 
We also note that while those who partake in contract lifecycle management do perform searches, potentially for an executive, that they are not professional searchers~\cite{verberne2019information,10.1145/3308774.3308799,wiggers_verberne_zwenne_vanloon_2022,RUSSELLROSE20181042} as this forms only a small part of their role in an organization and reducing the amount of searching needed would make them more effective otherwise (e.g., not having to search for renewal conditions).

We note that much of the data that might be collected in contract lifecycle systems are generated through extractive machine learning~\cite{leivaditi2020benchmark,hendrycks2021cuad,wang2023maud,10.1145/3209978.3210015} and other forms of classification (e.g., classifying contract type).
The exact extent of these capabilities was not discussed as part of these interviews but note that most dedicated CLM tools offer them to one extent or another.

\begin{figure}
\centering
\resizebox{0.95\columnwidth}{!}{
\begin{tabular}{c | p{9cm}}
Stage & Description \\ \hline
Drafting & This is where a contract is created for a particular purpose (e.g., employee hiring, NDA), usually based on an existing template in the organization. The template is specialized to the current needs and any substantive changes are usually approved by management. This stage is optional if the other party is drafting the agreement. \\ \hline
Negotiation & In this stage, the parties involved take turns editing and proposing changes to the agreement. Often, this results in red lines (e.g., proposed deletions and their substitutions) from one version of the agreement to another. Additionally, the proposed changes may be compared in a black line to a previous version to more easily understand deviations over time. Eventually, a final copy is proposed that is acceptable to the legal counsel of all parties involved. Parties may wish to save the individual versions for subsequent analysis.\\ \hline
Approval & A final approval stage is used to ensure that the proposed agreement does not place undue burden on either organization (e.g., agreeing to implausible development work, disproportionate support). \\ \hline
Signature & This is the formal end of contract negotiation. This is simply the signed copy of the approved agreement by all parties and is the source of truth for all contract questions going forward. Storing this version is critical for ensuring that contract terms are met. \\ \hline
Management & In this stage, the obligations and requirements imposed by the contract are enforced by both parties. Compliance with obligations can be problematic without good tracking, especially if every negotiation results in substantive differences from a template agreement. These obligations may also specify how and when renewal discussions should take place, which may result in amendments to the current contract or a completely new negotiation. \\ \hline
Reporting & A final component in a contract's life is its use in any reporting that might take place. This stage lets an organization undertake analysis on what they have agreed to and the variations in the agreements. Such analysis may be done to improve templates to reduce negotiation time but may also be done to understand risks or liabilities present in the contracts. \\ \hline
\end{tabular}}
\caption{\label{fig:lifecycle} A simplified overview of the contract lifecycle process. A contract will start in ``Drafting'' and move down.}
\end{figure}

\section{Methodology}
We conducted a set of semi-structured interviews held over Zoom. The interviews targeted 30 minutes in length but varied due to the time availability of the participants and their interest in continuing to talk. 

We recruited participants using convenience sampling performing a reach out on LinkedIn or via coworkers professional social circles. Our target group were members of Legal Operations teams who not only have an extensive legal background but are also commonly tasked with automating companies' legal workflows and otherwise managing contract and legal agreement repositories with Microsoft SharePoint (or a related system).
In total, 26 participants were recruited for this in-depth study. 

Over the course of the interviews, we asked participants about their approach to contract management, experience with different software products, and reasons for choosing their current framework. Prior to beginning the interview, participants were asked for their permission to record the call in order to facilitate subsequent analysis. Upon completing the interview process, one of the authors verified the automatically produced transcripts (via Otter.ai\footnote{\url{https://otter.ai}} or Zoom Transcripts) and analysed them using thematic analysis method.
In the case that a participant declined to be recorded or if there was a technical glitch, notes were taken and those were used in the thematic analysis. Thematic analysis revealed that the areas of contracts management with most room for improvement included lack of single shared storage, difficulty finding the right contracts at the right time, maintaining contracts after signing, and contract negotiation.

\section{Is that the contract and \textit{your} signature?}
The foundation of any contract management process is to store contracts and is so foundational to successful process that participant P141 opined that \textit{``first and foremost it needs to be a repository, one place for everything to live so that we are no longer looking at email, desktops, drives, Box, DropBox, etc, etc, etc. It's the first goal, the crawl \textlangle from crawl, walk, run metaphor\textrangle.''} Ironically, having a single storage is also identified as one of the major obstacles. For example, P148 has to deal with duplicated contracts, \textit{``unfortunately, we still have separated drives between Legal, Finance, and one which is more common to the whole firm. And everyone wants to have a copy of the signed agreement. ... For the moment, we duplicate the signed agreement in different drives.''} While P132-33, who works at a smaller company, finds that they are the bottleneck for contract storage and that they have to deal with this responsibility (\textit{``there's not a whole system of other people uploading anything if they have a contract that's executed. It winds up on my desk or in my email and I take care of it from there''}).

Additionally, effective contract management requires having all related documents linked in an transparent fashion to avoid, in the words of P129, \textit{``slogging through a list of files that don't seem to have any relationship to each other''}. Such related documents may include attachments, amendments, contract versions, etc. The situation further exacerbates by having companies acquired and renamed. P141 sums up: \textit{``ensuring that the children of the agreement make it in, amendments one, two, three, and four, an assignment from company A to their subsidiary, a change order three, SOWs, and then maybe an order form for whatever physical good might be a part of this whole thing. Keeping them together is goal one, and that's a really lofty goal.''}

Microsoft SharePoint comes to the rescue by creating a single space for contracts, as P127 points out, \textit{``SharePoint is great at organisation type tasks, or organising documents and storing documents.''} This is especially true for smaller companies with limited budgets, for example P148 recalls from their time working at a smaller startup that \textit{``we didn't want to invest directly in a dedicated tool. So we started using SharePoint.''}
Companies committed to Microsoft Suite will find added benefits in the familiarity of the interface and high level of integration with other Microsoft tools, speeding up the adoption process. P131 elaborates further: \textit{``[SharePoint] was something that people knew, people had a general understanding of, and were able to easily integrate it across those who needed it.''}

SharePoint's functionality is unfortunately lacking when is comes to setting up fine-grained access to sensitive documents. P129 describes their situation, \textit{``the challenge I face is you can't really create user groups with defined permissions. And instead it becomes ad-hoc, like this person can access this document. And that's really hard to stay on top of who get what when.''} Such lack of governance can lead to costly mistakes as experienced by P132-33, \textit{``we have a number of times run into situations where we have found a random contract that was signed by someone who should never have been authorised to sign it. And they agreed to something that caused the entire situation.''} Having more control over not just who can access a contract but who can modify it (or even sign it) means that organizations can be more certain about what they have agreed to, even if finding that information is not always easy.

\section{Seek Contractual Treasure}
\subsection{Avoiding Surprises}
Contract lifecycle management does not end with a contract being signed, on the contrary, it is only half the battle as expressed by P129, \textit{``\textlangle a contract\textrangle is a living breathing documents that just provides structure to a thing that you said you wanted to do. You should be monitoring what those key deadlines, those key deliverables that you expect, what your rights are, if something goes wrong it should be like a user guide or manual.''}

In tracking obligations, organizations ensure that promises made are promises kept and that there are not unexpected surprises (e.g., if a contract automatically renews). P143 compares a traditional storage repository to a file cabinet, \textit{``once you put something in a file cabinet, nobody ever goes in there. And that's what the repository was for us \textlangle ...\textrangle All of my engineers and operations would have to wait for something to go wrong. And then the customer would complain about it, and they would have to go manually find that document to say `Oh, what did we sign up to do?'' And it was always a day late and a dollar short.''}
This reactive and potentially slow response means that customer relationships can be damaged and could lead to breach of contract terms; neither option is particularly appealing. 

Many dedicated CLM tools provide such functionality out-of-the-box making it one of the selling points, as in the case of P134, \textit{``\textlangle CLM tool\textrangle is great at creating those kinds of scheduled workflows that will comb through everything and say `Oh, well, this agreement is about to expire in 6 months. Let's send somebody a notice. Who is the owner of this contract?'''}

In contrast, SharePoint provides users with the ability to create automated workflows using PowerAutomate, potentially alleviating such issues,  but such functionality remains untapped by many, including P145 who viewed SharePoint mostly as a repository, \textit{``the problem is that contract almost died more or less when you put it \textlangle in SharePoint\textrangle. You don't do follow ups, you don't know what happened next, which obligations are being fulfilled, or which are not.''} While others are aware of SharePoint's potential, they consider it clunky as expressed by P130, \textit{``you have to kind of go on the back end, manipulate some stuff and try to fit a square peg in a round hole a little bit.''} We will see this notion come up in discussing other aspects of SharePoint for contract management: the functionality does exist but is challenging to access and/or use effectively. 

\subsection{The Right Contract(s) at the Right Time} 
Searching for and through documents constitutes a large part of a legal professional's time. Our participants mentioned several cases in which locating the right documents was critical to job success which underscores the importance of search. For example, P127 found that speedy search was crucial in their use case,\textit{``you have your end of month, quarter and year, you need to be able to access information as quickly as possible''}. While P147 experienced a different situation requiring high recall, \textit{``if we were going into a litigation, finding all of the contracts became a really big chore''}. 

However, the volume of information usually stored in a contract repository stifles the process of finding documents effectively. Most participants elected to extract metadata from documents to improve their search process. Such metadata included customer name, contract sign date, date of contract renewal, and many others. Transitioning an unstructured contract text into a semi-structured format enabled sorting and filtering to be done. P145 summarises their experience as follows, \textit{``\textlangle~a challenging thing in SharePoint was\textrangle finding the information. For example, you have a ton of documents and then go find something in there. And the search engine of SharePoint is not very good. So we try to build some metadata in the documents just to classify them, so that we can try to find them easier.''} 

Microsoft Suite offers a tool called Syntex\footnote{\url{https://www.microsoft.com/en-us/microsoft-syntex}} for the purpose of extracting information from documents. Syntex provides its users with an ability to train a machine-learning model to find and extract concepts from unstructured document text. However, the old problem comes up again -- the functionality is not accessible enough to the users, so much so that many are unaware of its existence and say that SharePoint \textit{``has no AI''} (P144). Accordingly, people turn to dedicated CLM tools that put their information extraction capabilities front and centre despite the migration costs. 

\subsection{Surveying the Contract Landscape} 
Our participants expressed desire to take the information synthesis capabilities one step further and create various overviews of stored documents. 
One such overview might be a report reflecting a general idea of the kinds of contracts are being dealt with. These reports can be targeted to higher-level management, as P140 describes, \textit{``we need to consolidate information so that we can have a report of sorts for the executive team to review. The executives usually don't care about the individual contracts.''}. Alternatively, a structured overview can act as a more efficient way to present data, as in the case of P138, \textit{``I could also build a high level litigation report. Because now I am interested in collecting how many credit collection cases do we have? How many labour cases we have? But not entering the into the details of the case, just having the macro data for reporting purposes.''}
In some sense, these forms of reporting can allow individuals to demonstrate value or their meeting of key performance indicators.

Structured contract overview can also aid in a person's day-to-day job, summarising the most relevant information in one location, as expressed by P149, \textit{``the main page has all the infographics I want about all open agreements, or agreements that will terminate in the coming six months, three months, one month, etc. I can have data about the old agreements, the specific clause or specific date, or parties or whatever. So I can have a big picture with a lot of data in a single page.''}

Beyond just contract content and metadata, our participants were looking to get analytics that provide insight into their statistics around their contract management processes. Whether it is increasing transparency of Legal team's work (P146: \textit{``we want to be able to showcase what the whole team is doing to the rest of the organisation. And also honestly just show how constrained the legal team can be. You're not getting your NDA turned around in less than a day because there are some other matters. You can look at the queue and see where you're falling.''}), or analysing expenditures (\textit{``we want to be able to compare how much time was spent on outside council, versus how much our internal staff spent on this contract in this matter. And then be able to do the math on what's the relative rate of your in-house counsel versus your external counsel and where's the savings?''}), or looking to optimise the overall process (P132-33: \textit{``If I've got one provision that continues to be redlined, is it because of me? Is it because the law has changed? \textlangle...\textrangle Have I got one type of product that we're selling that's really causing a lot of friction in our sales process and our pipeline? Is it a contract issue? Is it a project issue?''}).
This additional view of the process allows legal operations teams to make more informed and better decisions about how they can better spend time and optimize their processes to ideally be more efficient and effective.

\section{Ain't you ever heard of negotiatin'?}
Although we did not specifically emphasise the topic of contract negotiation in our interviews, our participants organically brought it up on multiple occasions. The core discussion revolved around AI-aided contract generation to ease the workload on legal teams and empower non-legal professionals to produce low-risk straightforward contracts on their own (i.e., akin to a ``fill in the blanks'' contract template). For example, P146 paints the picture of their process, \textit{``we're doing some auto generation of documents using SharePoint and PowerAutomate that has really helped us speed up cycle times. \textlangle...\textrangle They are all low risk contracts at this point. \textlangle...\textrangle Now the NDAs go out without anyone on the Legal team touching them. \textlangle...\textrangle And we're a super small team, so this was definitely something that we did out of necessity. We just did not have the bandwidth to be handling hundreds of NDAs, hundreds of supply requests.''}
Such approaches highlight how beneficial such integrations can be for legal teams but highlight that SharePoint alone is not sufficient.
While P146 did not discuss the effort involved in setting up this integration, from our own experiences, we expect that there was non-trivial technical support in some capacity to set this up and get it running properly, depending on the exact complexity.

An ideal negotiation tool will also help optimise the process as time goes by, drawing its users’ attention to areas of contracts that repeatedly cause issues or grief. As P132-33 noted, \textit{``having a standard response to some of the regular pushback that we expect from our contracts is such a time saver.''}. Others express similar sentiment with P143 stating, \textit{``If we are going to consistently agree to do something, then let's just make that a part of the standard and not negotiate it each time we send a term.''}, and P141 echoing related way of thinking, \textit{``What are the most negotiated points within that contract? I would use that kind of intelligence to say well, if we're always agreeing to this change, because our starting position is a non-starter for everyone else, let's remove that bottleneck.''}
We also imagine that other forms of machine learning could be leveraged to suggest ``middle grounds'' between two different versions of a clause or such machine learning may just automatically negotiate the contract itself~\cite{oliver1996,bendraou2020}.
In either case, ensuring that incorporating this functionality into existing processes is as seamless as possible would greatly foster adoption. 

Notably, in our discussions of the negotiation process people also mentioned that legal professionals are partial to using Microsoft Word to conduct negotiations and moving the industry into a different tool for this purpose will prove to be an arduous, or possibly Sisyphean, task to undertake. 
Tools like SharePoint will have a natural advantage in facilitating the use of Microsoft Word (for obvious reasons) but it is still not an easy task to ensure desired functionality is consistently available across all different possible access points (i.e., functional equivalence of add-ons in SharePoint and Word).


\section{Conclusion: We're in a tight spot}
Microsoft SharePoint has the potential to become popular as a contract management tool. The familiar ecosystem, heavy integration, and attractive pricing model make it a go-to choice for smaller teams who are looking to begin their journey on the road fraught with peril towards the treasure that is an effective contract organisation system. On paper, SharePoint provides its users with many capabilities, however, in reality, using many of its functions requires a certain level of tech-savvyness and a big time commitment. Suffice it to say that a lot of the legal and legal operations professionals responsible for contract lifecycle management may not be in a position to use a tool like this. Anecdotally, in the process of working on this project, we asked one of the non-tech people at our company to create a PowerAutomate flow which ultimately was a task that did not end in success. Conversely, stand-alone CLM providers offer out-of-the-box functionality that is easily discoverable and designed with a contract management process in mind. However, their pricing far exceeds that of SharePoint. Additionally, the process of transitioning to a new tool is costly as well, while its adoption within the company may be slow. 

All in all, neither option seems to have a clear advantage over the other and the companies looking to establish a contract\--management process should choose according to their situation and preferences.


\bibliographystyle{ACM-Reference-Format}
\bibliography{sample-base}


\begin{thebibliography}{22}


\ifx \showCODEN    \undefined \def \showCODEN     #1{\unskip}     \fi
\ifx \showDOI      \undefined \def \showDOI       #1{#1}\fi
\ifx \showISBNx    \undefined \def \showISBNx     #1{\unskip}     \fi
\ifx \showISBNxiii \undefined \def \showISBNxiii  #1{\unskip}     \fi
\ifx \showISSN     \undefined \def \showISSN      #1{\unskip}     \fi
\ifx \showLCCN     \undefined \def \showLCCN      #1{\unskip}     \fi
\ifx \shownote     \undefined \def \shownote      #1{#1}          \fi
\ifx \showarticletitle \undefined \def \showarticletitle #1{#1}   \fi
\ifx \showURL      \undefined \def \showURL       {\relax}        \fi
\providecommand\bibfield[2]{#2}
\providecommand\bibinfo[2]{#2}
\providecommand\natexlab[1]{#1}
\providecommand\showeprint[2][]{arXiv:#2}

\bibitem[Arguello et~al\mbox{.}(2021)]%
        {10.1145/3406522.3446021}
\bibfield{author}{\bibinfo{person}{Jaime Arguello}, \bibinfo{person}{Adam Ferguson}, \bibinfo{person}{Emery Fine}, \bibinfo{person}{Bhaskar Mitra}, \bibinfo{person}{Hamed Zamani}, {and} \bibinfo{person}{Fernando Diaz}.} \bibinfo{year}{2021}\natexlab{}.
\newblock \showarticletitle{Tip of the Tongue Known-Item Retrieval: A Case Study in Movie Identification}. In \bibinfo{booktitle}{\emph{Proceedings of the 2021 Conference on Human Information Interaction and Retrieval}} (Canberra ACT, Australia) \emph{(\bibinfo{series}{CHIIR '21})}. \bibinfo{publisher}{Association for Computing Machinery}, \bibinfo{address}{New York, NY, USA}, \bibinfo{pages}{5–14}.
\newblock
\showISBNx{9781450380553}
\urldef\tempurl%
\url{https://doi.org/10.1145/3406522.3446021}
\showDOI{\tempurl}


\bibitem[Broder(2002)]%
        {10.1145/792550.792552}
\bibfield{author}{\bibinfo{person}{Andrei Broder}.} \bibinfo{year}{2002}\natexlab{}.
\newblock \showarticletitle{A Taxonomy of Web Search}.
\newblock \bibinfo{journal}{\emph{SIGIR Forum}} \bibinfo{volume}{36}, \bibinfo{number}{2} (\bibinfo{date}{sep} \bibinfo{year}{2002}), \bibinfo{pages}{3–10}.
\newblock
\showISSN{0163-5840}
\urldef\tempurl%
\url{https://doi.org/10.1145/792550.792552}
\showDOI{\tempurl}


\bibitem[Dumais et~al\mbox{.}(2003)]%
        {dumais2003}
\bibfield{author}{\bibinfo{person}{Susan Dumais}, \bibinfo{person}{Edward Cutrell}, \bibinfo{person}{JJ Cadiz}, \bibinfo{person}{Gavin Jancke}, \bibinfo{person}{Raman Sarin}, {and} \bibinfo{person}{Daniel~C. Robbins}.} \bibinfo{year}{2003}\natexlab{}.
\newblock \showarticletitle{Stuff I've Seen: A System for Personal Information Retrieval and Re-Use}. In \bibinfo{booktitle}{\emph{Proceedings of the 26th Annual International ACM SIGIR Conference on Research and Development in Informaion Retrieval}} (Toronto, Canada) \emph{(\bibinfo{series}{SIGIR '03})}. \bibinfo{publisher}{Association for Computing Machinery}, \bibinfo{address}{New York, NY, USA}, \bibinfo{pages}{72–79}.
\newblock
\showISBNx{1581136463}
\urldef\tempurl%
\url{https://doi.org/10.1145/860435.860451}
\showDOI{\tempurl}


\bibitem[Hawking(2004)]%
        {hawking2004challenges}
\bibfield{author}{\bibinfo{person}{David Hawking}.} \bibinfo{year}{2004}\natexlab{}.
\newblock \showarticletitle{Challenges in Enterprise Search.}. In \bibinfo{booktitle}{\emph{ADC}}, Vol.~\bibinfo{volume}{4}. \bibinfo{pages}{15--24}.
\newblock


\bibitem[Hendrycks et~al\mbox{.}(2021)]%
        {hendrycks2021cuad}
\bibfield{author}{\bibinfo{person}{Dan Hendrycks}, \bibinfo{person}{Collin Burns}, \bibinfo{person}{Anya Chen}, {and} \bibinfo{person}{Spencer Ball}.} \bibinfo{year}{2021}\natexlab{}.
\newblock \bibinfo{title}{CUAD: An Expert-Annotated NLP Dataset for Legal Contract Review}.
\newblock
\newblock
\showeprint[arxiv]{2103.06268}~[cs.CL]


\bibitem[Insights(2023)]%
        {clm-market}
\bibfield{author}{\bibinfo{person}{Fortune~Business Insights}.} \bibinfo{year}{2023}\natexlab{}.
\newblock \bibinfo{title}{Contract Lifecycle Management Market Report}.
\newblock \bibinfo{howpublished}{\url{https://www.fortunebusinessinsights.com/contract-lifecycle-management-clm-solution-market-106472}}.
\newblock


\bibitem[Kruschwitz(2023)]%
        {kruschwitz20233}
\bibfield{author}{\bibinfo{person}{Udo Kruschwitz}.} \bibinfo{year}{2023}\natexlab{}.
\newblock \showarticletitle{3.12 (Aspects of) Enterprise Search}.
\newblock \bibinfo{journal}{\emph{Frontiers of Information Access Experimentation for Research and Education}} (\bibinfo{year}{2023}), \bibinfo{pages}{12}.
\newblock


\bibitem[Kruschwitz and Hull(2017)]%
        {INR-053}
\bibfield{author}{\bibinfo{person}{Udo Kruschwitz} {and} \bibinfo{person}{Charlie Hull}.} \bibinfo{year}{2017}\natexlab{}.
\newblock \showarticletitle{Searching the Enterprise}.
\newblock \bibinfo{journal}{\emph{Foundations and Trends® in Information Retrieval}} \bibinfo{volume}{11}, \bibinfo{number}{1} (\bibinfo{year}{2017}), \bibinfo{pages}{1--142}.
\newblock
\showISSN{1554-0669}
\urldef\tempurl%
\url{https://doi.org/10.1561/1500000053}
\showDOI{\tempurl}


\bibitem[Leivaditi et~al\mbox{.}(2020)]%
        {leivaditi2020benchmark}
\bibfield{author}{\bibinfo{person}{Spyretta Leivaditi}, \bibinfo{person}{Julien Rossi}, {and} \bibinfo{person}{Evangelos Kanoulas}.} \bibinfo{year}{2020}\natexlab{}.
\newblock \bibinfo{title}{A Benchmark for Lease Contract Review}.
\newblock
\newblock
\showeprint[arxiv]{2010.10386}~[cs.IR]


\bibitem[Meier and Elsweiler(2018)]%
        {10.1145/3176349.3176392}
\bibfield{author}{\bibinfo{person}{Florian Meier} {and} \bibinfo{person}{David Elsweiler}.} \bibinfo{year}{2018}\natexlab{}.
\newblock \showarticletitle{Other Times It\'{z}s Just Strolling Back Through My Timeline: Investigating Re-Finding Behaviour on Twitter and Its Motivations}. In \bibinfo{booktitle}{\emph{Proceedings of the 2018 Conference on Human Information Interaction \& Retrieval}} (New Brunswick, NJ, USA) \emph{(\bibinfo{series}{CHIIR '18})}. \bibinfo{publisher}{Association for Computing Machinery}, \bibinfo{address}{New York, NY, USA}, \bibinfo{pages}{130–139}.
\newblock
\showISBNx{9781450349253}
\urldef\tempurl%
\url{https://doi.org/10.1145/3176349.3176392}
\showDOI{\tempurl}


\bibitem[Mukherjee and Mao(2004)]%
        {mukherjee2004}
\bibfield{author}{\bibinfo{person}{Rajat Mukherjee} {and} \bibinfo{person}{Jianchang Mao}.} \bibinfo{year}{2004}\natexlab{}.
\newblock \showarticletitle{Enterprise Search: Tough Stuff: Why is It That Searching an Intranet is so Much Harder than Searching the Web?}
\newblock \bibinfo{journal}{\emph{Queue}} \bibinfo{volume}{2}, \bibinfo{number}{2} (\bibinfo{date}{apr} \bibinfo{year}{2004}), \bibinfo{pages}{36–46}.
\newblock
\showISSN{1542-7730}
\urldef\tempurl%
\url{https://doi.org/10.1145/988392.988406}
\showDOI{\tempurl}


\bibitem[Oliver(1996)]%
        {oliver1996}
\bibfield{author}{\bibinfo{person}{Jim Oliver}.} \bibinfo{year}{1996}\natexlab{}.
\newblock \showarticletitle{A Machine-Learning Approach to Automated Negotiation and Prospects for Electronic Commerce}.
\newblock \bibinfo{journal}{\emph{Journal of Management Information Systems}}  \bibinfo{volume}{13} (\bibinfo{date}{12} \bibinfo{year}{1996}), \bibinfo{pages}{83--112}.
\newblock
\urldef\tempurl%
\url{https://doi.org/10.1080/07421222.1996.11518135}
\showDOI{\tempurl}


\bibitem[Roegiest et~al\mbox{.}(2018)]%
        {10.1145/3209978.3210015}
\bibfield{author}{\bibinfo{person}{Adam Roegiest}, \bibinfo{person}{Alexander~K. Hudek}, {and} \bibinfo{person}{Anne McNulty}.} \bibinfo{year}{2018}\natexlab{}.
\newblock \showarticletitle{A Dataset and an Examination of Identifying Passages for Due Diligence}. In \bibinfo{booktitle}{\emph{The 41st International ACM SIGIR Conference on Research \& Development in Information Retrieval}} (Ann Arbor, MI, USA) \emph{(\bibinfo{series}{SIGIR '18})}. \bibinfo{publisher}{Association for Computing Machinery}, \bibinfo{address}{New York, NY, USA}, \bibinfo{pages}{465–474}.
\newblock
\showISBNx{9781450356572}
\urldef\tempurl%
\url{https://doi.org/10.1145/3209978.3210015}
\showDOI{\tempurl}


\bibitem[Russell-Rose et~al\mbox{.}(2018)]%
        {RUSSELLROSE20181042}
\bibfield{author}{\bibinfo{person}{Tony Russell-Rose}, \bibinfo{person}{Jon Chamberlain}, {and} \bibinfo{person}{Leif Azzopardi}.} \bibinfo{year}{2018}\natexlab{}.
\newblock \showarticletitle{Information retrieval in the workplace: A comparison of professional search practices}.
\newblock \bibinfo{journal}{\emph{Information Processing \& Management}} \bibinfo{volume}{54}, \bibinfo{number}{6} (\bibinfo{year}{2018}), \bibinfo{pages}{1042--1057}.
\newblock
\showISSN{0306-4573}
\urldef\tempurl%
\url{https://doi.org/10.1016/j.ipm.2018.07.003}
\showDOI{\tempurl}


\bibitem[Russell-Rose et~al\mbox{.}(2011)]%
        {russell2011taxonomy}
\bibfield{author}{\bibinfo{person}{Tony Russell-Rose}, \bibinfo{person}{Joe Lamantia}, {and} \bibinfo{person}{Mark Burrell}.} \bibinfo{year}{2011}\natexlab{}.
\newblock \showarticletitle{A Taxonomy of Enterprise Search.}. In \bibinfo{booktitle}{\emph{EuroHCIR}}. \bibinfo{pages}{15--18}.
\newblock


\bibitem[Sommers et~al\mbox{.}(2023)]%
        {gartner}
\bibfield{author}{\bibinfo{person}{Kaitlynn Sommers}, \bibinfo{person}{Lynne Phelan}, {and} \bibinfo{person}{Kerrie McDonald}.} \bibinfo{year}{2023}\natexlab{}.
\newblock \bibinfo{title}{Gartner Report: Magic Quadrant for Contract Life Cycle Management}.
\newblock
\newblock


\bibitem[Stocker et~al\mbox{.}(2014)]%
        {stocker2014}
\bibfield{author}{\bibinfo{person}{Alexander Stocker}, \bibinfo{person}{Markus Zoier}, \bibinfo{person}{Selver Softic}, \bibinfo{person}{Stefan Paschke}, \bibinfo{person}{Heimo Bischofter}, {and} \bibinfo{person}{Roman Kern}.} \bibinfo{year}{2014}\natexlab{}.
\newblock \showarticletitle{Is Enterprise Search Useful at All? Lessons Learned from Studying User Behavior}. In \bibinfo{booktitle}{\emph{Proceedings of the 14th International Conference on Knowledge Technologies and Data-Driven Business}} (Graz, Austria) \emph{(\bibinfo{series}{i-KNOW '14})}. \bibinfo{publisher}{Association for Computing Machinery}, \bibinfo{address}{New York, NY, USA}, Article \bibinfo{articleno}{22}, \bibinfo{numpages}{8}~pages.
\newblock
\showISBNx{9781450327695}
\urldef\tempurl%
\url{https://doi.org/10.1145/2637748.2638425}
\showDOI{\tempurl}


\bibitem[Verberne et~al\mbox{.}(2019a)]%
        {10.1145/3308774.3308799}
\bibfield{author}{\bibinfo{person}{Suzan Verberne}, \bibinfo{person}{Jiyin He}, \bibinfo{person}{Udo Kruschwitz}, \bibinfo{person}{Gineke Wiggers}, \bibinfo{person}{Birger Larsen}, \bibinfo{person}{Tony Russell-Rose}, {and} \bibinfo{person}{Arjen~P. de Vries}.} \bibinfo{year}{2019}\natexlab{a}.
\newblock \showarticletitle{First International Workshop on Professional Search}.
\newblock \bibinfo{journal}{\emph{SIGIR Forum}} \bibinfo{volume}{52}, \bibinfo{number}{2} (\bibinfo{date}{jan} \bibinfo{year}{2019}), \bibinfo{pages}{153–162}.
\newblock
\showISSN{0163-5840}
\urldef\tempurl%
\url{https://doi.org/10.1145/3308774.3308799}
\showDOI{\tempurl}


\bibitem[Verberne et~al\mbox{.}(2019b)]%
        {verberne2019information}
\bibfield{author}{\bibinfo{person}{Suzan Verberne}, \bibinfo{person}{Jiyin He}, \bibinfo{person}{Gineke Wiggers}, \bibinfo{person}{Tony Russell-Rose}, \bibinfo{person}{Udo Kruschwitz}, {and} \bibinfo{person}{Arjen~P. de Vries}.} \bibinfo{year}{2019}\natexlab{b}.
\newblock \bibinfo{title}{Information search in a professional context - exploring a collection of professional search tasks}.
\newblock
\newblock
\showeprint[arxiv]{1905.04577}~[cs.IR]


\bibitem[Wang et~al\mbox{.}(2023)]%
        {wang2023maud}
\bibfield{author}{\bibinfo{person}{Steven~H. Wang}, \bibinfo{person}{Antoine Scardigli}, \bibinfo{person}{Leonard Tang}, \bibinfo{person}{Wei Chen}, \bibinfo{person}{Dimitry Levkin}, \bibinfo{person}{Anya Chen}, \bibinfo{person}{Spencer Ball}, \bibinfo{person}{Thomas Woodside}, \bibinfo{person}{Oliver Zhang}, {and} \bibinfo{person}{Dan Hendrycks}.} \bibinfo{year}{2023}\natexlab{}.
\newblock \bibinfo{title}{MAUD: An Expert-Annotated Legal NLP Dataset for Merger Agreement Understanding}.
\newblock
\newblock
\showeprint[arxiv]{2301.00876}~[cs.CL]


\bibitem[Wiggers et~al\mbox{.}(2022)]%
        {wiggers_verberne_zwenne_vanloon_2022}
\bibfield{author}{\bibinfo{person}{Gineke Wiggers}, \bibinfo{person}{Suzan Verberne}, \bibinfo{person}{Gerrit-Jan Zwenne}, {and} \bibinfo{person}{Wouter Van~Loon}.} \bibinfo{year}{2022}\natexlab{}.
\newblock \showarticletitle{Exploration of Domain Relevance by Legal Professionals in Information Retrieval Systems}.
\newblock \bibinfo{journal}{\emph{Legal Information Management}} \bibinfo{volume}{22}, \bibinfo{number}{1} (\bibinfo{year}{2022}), \bibinfo{pages}{49–67}.
\newblock
\urldef\tempurl%
\url{https://doi.org/10.1017/S1472669622000093}
\showDOI{\tempurl}


\bibitem[Youssef et~al\mbox{.}(2020)]%
        {bendraou2020}
\bibfield{author}{\bibinfo{person}{Bendraou Youssef}, \bibinfo{person}{Nathan Ramoly}, {and} \bibinfo{person}{Baba Bamba}.} \bibinfo{year}{2020}\natexlab{}.
\newblock \showarticletitle{Toward a Generic AutoML-Based Assistant for Contracts Negotiation}.
\newblock


\end{thebibliography}

\end{document}